\documentstyle[11pt]{article}

\begin{document}
\begin{center}
{\Large {\bf NONLINEAR GRAVITON AS A LIMIT OF {\it sl}(N;C) CHIRAL 
FIELDS AS N$\rightarrow\infty$}}
\vspace{4ex}

\bf Maciej Przanowski \rm
\footnote[2]{e-mail: mprzan@ck-sg.p.lodz.pl}\\
\vspace{3 ex}

Institute of Physics, Technical University of \L\'{o}d\'{z},\\
W\'{o}lcza\'{n}ska 219. 93-005 \L\'{o}d\'{z}, Poland\\
\vspace{1 ex}

Department of Theoretical Physics, University of \L\'{o}d\'{z},\\
Pomorska 149/153.  90-236 \L\'{o}d\'{z}, Poland\\
\vspace{3 ex}

\bf Sebastian Forma\'{n}ski \rm
\footnote[3]{e-mail: sforman@ck-sg.p.lodz.pl} and 
\bf Francisco J. Turrubiates \rm
\footnote[8]{Permanent address:Departamento de F\'{i}sica,
Centro de Investigaci\'{o}n y de Estudios Avanzados del IPN,
Apartado Postal 14-740, 07000 M\'{e}xico D.F., M\'{e}xico\\
e-mail: fturrub@fis.cinvestav.mx}\\

\vspace{2 ex}

Institute of Physics, Technical University of \L\'{o}d\'{z},\\
W\'{o}lcza\'{n}ska 219. 93-005 \L\'{o}d\'{z}, Poland

\end{center}

\vspace{3 ex}

\section*{Abstract}
An example of a sequence of the {\it sl}(N;C) chiral fields, for N$\geq 2$, tending 
to the complex heavenly metric (nonlinear graviton) of the type 
$[4]\times [-]$ when N$\rightarrow\infty$ is given.

\vspace {3ex}

\noindent Keywords:

\noindent Self-dual gravity, chiral fields, infinite-dimensional Lie algebras.

\vspace {3ex}

\section{Introduction}
Ward \cite{1}, Park \cite{2}, Strachan \cite{3} and Husain \cite{4} were 
the first who observed the close relation between the principal chiral 
model and self-dual gravity. Then in \cite{5,6,7} this relation has been 
analysed from several points of view. Briefly speaking, self-dual gravity 
on a real 4-dimensional manifold endowed with metric of signature 
$(++ - - )$ may be locally considered to be the principal chiral model 
on $ V\subset R^2$ for the real Poisson algebra of a 2-surface $\Sigma^2$.
Assuming that  $\Sigma^2$ is the 2-torus $T^2$ one can construct the 
isomorphism between the real Poisson algebra of $T^2$ and the 
{\it su}$(\infty )$ algebra. Similarly, the complex Poisson algebra of 
$T^2$ is isomorphic to the {\it sl}$(\infty ;C)$ algebra.

Thus the natural question arises: ''Can one construct a sequence of the 
{\it su}(N) chiral fields, for $N=2,3,\dots ,$ tending to a curved 
space in a limit". This question was first stated by Ward \cite{1}.
In previous works \cite{6,7} some approach to {\it Ward's question} has 
been proposed.

The aim of the present paper is to  give an example of a sequence of 
{\it sl}$(N,C)$ chiral fields for $N\geq 2$, tending to a curved heavenly 
space-time for $N\rightarrow\infty$.

Our paper is organized as follows. In Section 2 we recall the Husain-Park 
heavenly equation and its Moyal deformation. Section 3 is devoted to  
{\it Ward's question. }Here we present an approach to this question given 
in the previous works \cite{6,7}. Finally, in Section 4 we give an example 
of the {\it sl}$(N;C)$ chiral fields on $R^2$, or $C^2$, for $N=2,3,\dots,$ 
tending to the complex heavenly metric on a real space time, or on the 
complex space-time, respectively, for $N\rightarrow\infty$. This 
heavenly metric (nonlinear graviton) appears to be  of the type $[4]\times [-]$.

\section{Husain-Park heavenly equation}
V.Husain \cite{4} has shown that the Ashtekar-Jacobson-Smolin equations 
describing the heavenly metric can be reduced to the following equation
	\begin{equation}
	\label{2.1}
	\partial^{2}_{x}\Theta_{0}+\partial^{2}_{y}\Theta_{0}
	+ \{\partial_{x}\Theta_{0}\, , \partial_{y}\Theta_{0}\}_{{\cal P}}=0
	\end{equation}
where $\Theta_{0}=\Theta_{0}(x,y,p,q)$ and $\{\cdot , \cdot \}_{{\cal P}}$ 
stands for the Poisson bracket i.e.,
	\begin{displaymath}
	 \{\partial_{x}\Theta_{0}\, , \partial_{y}\Theta_{0}\}_{{\cal P}}=
	\partial_{q}( \partial_{x}\Theta_{0}) \partial_{p}( \partial_{y}\Theta_{0})-
	\partial_{q}( \partial_{y}\Theta_{0}) \partial_{p}( \partial_{x}\Theta_{0})
	\end{displaymath}
Then the heavenly metric $ds^2$ reads
	\begin{eqnarray}
	\label{2.2}
	ds^{2}= dx ( \partial_{p} \partial_{x} \Theta_{0} dp + 
	\partial_{q} \partial_{x} \Theta_{0} dq) +
	dy ( \partial_{p}\partial_{y} \Theta_{0}dp +
	\partial_{q}\partial_{y} \Theta_{0} dq) \nonumber\\
	-\, \, \frac{1}{ \{ \partial_{x} \Theta_{0}\, , \partial_{y} \Theta_{0}\}_{{\cal P}}}
	[(\partial_{p} \partial_{x} \Theta_{0} dp + 
	\partial_{q}\partial_{x} \Theta_{0} dq)^2 + 
	(\partial_{p}\partial_{y} \Theta_{0} dp + 
	\partial_{q}\partial_{y} \Theta_{0} dq)^2] 
	\end{eqnarray}
Observe that in the works \cite{4,5,6} there are some mistakes in sign in 
formulae for $ds^2$.

Equation (\ref{2.1}) has been also found by Park \cite{2}. (However, in his 
paper \cite{2} Park was not aware that (\ref{2.1}) defined the metric of a 
heavenly space-time). We call Eq. (\ref{2.1}) the {\it Husain-Park heavenly 
equation (H-P equation).}

As it has been pointed out in the previous papers \cite{5,6,7} it seems to 
be very fruitful to deal with the {\it Moyal deformation of the H-P equation} 
	\begin{eqnarray}
	\label{2.3}
	& \partial^{2}_{x}\Theta+\partial^{2}_{y}\Theta
	+ \{\partial_{x}\Theta\, , \partial_{y}\Theta\}_{{\cal M}}=0 & \nonumber \\
	& \Theta = \Theta (\hbar; x,y,p,q) \, , & 
	\end{eqnarray}
where $\{\cdot\, ,\cdot\}_{{\cal M}}$ stands for the Moyal bracket i.e.,
	\begin{eqnarray}
	\label{2.4}
	\{f_1\, , f_2\}_{{\cal M}} & := &
	f_1\, \frac{2}{\hbar}\sin(\frac{\hbar}{2} \stackrel{\leftrightarrow}{{\cal P}})f_2
	\nonumber\\
	 \stackrel{\leftrightarrow}{{\cal P}} & := & 
	 \stackrel{\leftarrow}{\frac{\partial}{\partial q}}
	 \stackrel{\rightarrow}{\frac{\partial }{ \partial p}}-
	 \stackrel{\leftarrow}{\frac{\partial}{\partial p}}
	 \stackrel{\rightarrow}{\frac{\partial}{\partial q}}
	\end{eqnarray}
$\hbar$ is the deformation parameter. 

It is evident that if $f_1$ and $f_2$ are independent of $\hbar$ then
	\begin{equation}
	\label{2.5}
	\lim_{\hbar \rightarrow 0}\{f_1\, , f_2\}_{{\cal M}}=
	\{f_1\, , f_2 \}_{{\cal P}}
	\end{equation}	
Let $\Theta = \Theta(\hbar; x,y,p,q)$ be an analytic in $\hbar$ solution of 
Eq. (\ref{2.3}) i.e., 
	\begin{equation}
	\label{2.6}
	\Theta = \sum_{n=0}^{\infty} \hbar^{n}\Theta_{n}\hspace{3 ex}
	\Theta_{n} =\Theta_{}(x,y,p,q)
	\end{equation}
Consequently one quickly finds that the function $\Theta_{0}$ satisfies the 
H-P equation (\ref{2.1}).

In the present paper we deal with solutions of Eq.(\ref{2.3}) on $V\times T^2$
 i.e., $(x,y)\in V$ and $(p,q)\in T^2$, where $V\subset R^2$ and $T^2$ is the 
2-torus of periods $2 \pi$ and $2\pi$. Thus $\Theta$ can be written in the 
following form
	\begin{equation}
	\label{2.7}
	\Theta = \sum_{(m_1,m_2)\in Z\times Z}\Theta_{(m_1,m_2)}(\hbar;x,y)
	\exp[i(m_1 p+ m_2 q)]
	\end{equation}
It is an easy matter to show that from (\ref{2.4}) one gets
	\begin{eqnarray}
	\label{2.8}	
	& \{E_{(m_1,m_2)}\, , E_{(n_1,n_2)} \}_{{\cal M}} = 
	\frac{2}{\hbar}\sin[\frac{\hbar}{2}(m_1 n_2-m_2 n_1)] E_{(m_1+n_1,m_2+n_2)} &
	\nonumber\\
	& (m_1,m_2),(n_1,n_2)\in Z\times Z & 
	\end{eqnarray}
where
	\begin{equation}
	\label{2.9}
	E_{(m_1,m_2)}:=\exp[i(m_1 p+ m_2 q)]
	\end{equation}

\section{The {\it sl}(N;C) chiral fields}
\setcounter{equation}{0}
Consider the {\it sl}(N;C) principal chiral model. The field equations read now
	\begin{equation}
	\label{3.1}
	\partial_{x} A_{y} - \partial_{y} A_{x} + [\; A_{x}\; , A_{y}\;]=0
	\end{equation}
	\begin{equation}
	\label{3.2}
	\partial_{x} A_{x} + \partial_{y} A_{y}=0
	\end{equation}
where $A_{x}=A_{x}(x,y)$ and $A_{y}=A_{y}(x,y)$ are {\it sl}(N;C)-valued 
functions on $V\subset R^2$. From (\ref{3.2}) one infers that there exists an 
{\it sl}(N;C)-valued function $\theta = \theta (x,y)$ such that
	\begin{equation}
	\label{3.3}
	A_{x}= -\partial_{y}\theta\hspace{4 em} {\rm and} \hspace{4 em}
	A_{y}= \partial_{x}\theta
	\end{equation}
Inserting (\ref{3.3}) into (\ref{3.1}) we get 
	\begin{equation}
	\label{3.4}
	\partial^{2}_{x}\theta + \partial^{2}_{y} \theta +
	[\; \partial_{x}\theta\; , \partial_{y}\theta\; ]=0
	\end{equation}
(Compare with \cite{5,6,7}). In the classical field theory Eq.(\ref{3.4})
can be equivalently considered to be the field equation of the {\it sl}(N;C)
principal chiral model.

To understand the relation between self-dual gravity and {\it sl}(N;C) 
principal chiral model we use some results of a distinguished paper by
Fairlie, Fletcher and Zachos \cite{8}. In particular in \cite{8} the basis
of the {\it su}(N) algebra has been considered which plays a crucial role in our 
construction. To define this basis consider $N\times N$ matrices
	\begin{eqnarray*}
	S:=\sqrt{\omega}
	\left(
	\begin{array}{ccccc}
	1&0&0&\ldots&0\\
	0&\omega&0&\ldots&0\\
	0&0&\omega^{2}&\ldots&0\\
	.&.&.&\ldots&.\\
	0&0&0&\ldots&\omega^{N-1}
	\end{array} \right)
	\end{eqnarray*}
	
	\begin{displaymath}
	\omega :=\exp(\frac{2\pi i}{N})\: , \hspace{3 em} 
	\sqrt{\omega}=\exp(\frac{\pi i}{N})
	\end{displaymath}
	
	\begin{eqnarray}
	\label{3.5}
	T:=\left(
	\begin{array}{ccccc}
	0&1&0&\ldots&0\\
	0&0&1&\ldots&0\\
	.&.&.&.&.\\
	0&0&0&\ldots&1\\
	-1&0&0&\ldots&0
	\end{array} \right)
	\end{eqnarray}

	\begin{displaymath}
	S^N=T^N=-1 \: , \hspace{4 em} T\cdot S=\omega S\cdot T
	\end{displaymath}
Then one defines
	\begin{equation}
	L_{(m_1,m_2)}:=\frac{iN}{2\pi} \omega^{\frac{m_1 m_2}{2}} S^{m_1}T^{m_2}; \hspace{4 em} 	(m_1,m_2)\in Z \times Z
	\label{3.6}
	\end{equation}
It is an easy matter to show that
	\begin{equation}
	\label{3.7}
	L_{(m_1,m_2)+ N(r_1,r_2)}= (-1)^{(m_1+1)r_2+(m_2+1)r_1+Nr_1 r_2} L_{(m_1,m_2)}
	\end{equation}
and	
	\begin{equation}
	[\;L_{(m_1,m_2)}\; , L_{(n_1,n_2)}\;]=
	\frac{N}{\pi}\sin[\frac{\pi}{N}(m_1n_2-m_2n_1)]
	L_{(m_1+n_1,m_2+n_2)}
	\label{3.8}
	\end{equation}
As it has been proved in \cite{8} $N^2 -1$ matrices $L_{(\mu_1 ,\mu_2 )}$ , 
$0\leq\mu_{1}\leq N-1$, $0\leq\mu_{2}\leq N-1$ and $(\mu_{1}, \mu_{2})\neq (0,0)$,
span the {\it su}(N) algebra (and of course the {\it sl}(N;C) algebra).
In what follows we assume that the Greek indices $(\mu_{1}, \mu_{2})$, 
$(\nu_{1}, \nu_{2}),\dots ,$ etc. satisfy the above written conditions). 

Comparing (\ref{2.8}) and (\ref{2.9}) with (\ref{3.8}), employing also (\ref{3.7})
one finds that the linear extension of the mapping
	\begin{eqnarray}
	\begin{array}{ccc}
	E_{(\mu_1 ,\mu_2 )+N(r_1 ,r_2 )}&\longmapsto &L_{(\mu_1 ,\mu_2 )+N(r_1 ,r_2 )}=
	(-1)^{(\mu_1+1)r_2+(\mu_2+1)r_1+Nr_1 r_2} L_{(\mu_1 ,\mu_2 )}\\
	 & & \\	
	E_{N(r_1 ,r_2 )}&\longmapsto &0
	\end{array}
	\label{3.9}
	\end{eqnarray}

	\begin{displaymath}
	0\leq\mu_{1}\leq N-1\hspace{0.5 em},\hspace{1 em} 0\leq\mu_{2}\leq N-1
	\hspace{1 em} and \hspace{1 em} (\mu_{1}, \mu_{2})\neq (0,0)
	\end{displaymath}
is a {\it Lie algebra homomorphism of the complex Moyal bracket algebra on} $T^2$ 
{\it with} $\hbar=\frac{2\pi}{N}$ {\it onto the Lie algebra sl}(N;C). Denote this 
homomorphism by $\Psi$. Let $\Theta=\Theta (\hbar; x,y,p,q)$ be some solution of the 
Moyal deformation of the H-P equation (\ref{2.3}) on $V \times T^2$. Then we
write $\Theta$ in the form (\ref{2.7}) and we put $\hbar=\frac{2\pi}{N}$. Define
	\begin{eqnarray}
	\label{3.10}
	\theta & = & \theta(N ;x,y):=  \Psi(\Theta(\frac{2\pi}{N} ;x,y,p,q))
	\nonumber\\	
	& = & \sum_{(\mu_{1},\mu_{2})} \{ \sum_{(r_{1},r_{2})\in Z\times Z}
	(-1)^{(\mu_1+1)r_2+(\mu_2+1)r_1+Nr_1 r_2} 
	\Theta_{(\mu_{1},\mu_{2})+N(r_{1},r_{2})}(\frac{2\pi}{N};x,y)\}
	L_{(\mu_1 ,\mu_2 )}\nonumber\\
	\end{eqnarray}
As $\Psi$ is a Lie algebra homomorphism the {\it sl}(N;C)-valued function
	\begin{displaymath}
	\theta = \theta(N ;x,y)= 
	\sum_{(\mu_1 ,\mu_2 )}\theta_{(\mu_1 ,\mu_2 )}(N;x,y) L_{(\mu_1 ,\mu_2 )}
	\end{displaymath}
	\begin{equation}
	\label{3.11}
	\theta_{(\mu_1 ,\mu_2 )}= \theta_{(\mu_1 ,\mu_2 )}(N;x,y):=
	\sum_{(r_{1},r_{2})\in Z\times Z}
	(-1)^{(\mu_1+1)r_2+(\mu_2+1)r_1+Nr_1 r_2} 
	\Theta_{(\mu_{1},\mu_{2})+N(r_{1},r_{2})}(\frac{2\pi}{N} ;x,y)
	\end{equation}
	\begin{displaymath}
	0\leq\mu_{1}\leq N-1\hspace{0.5 em},\hspace{1 em} 0\leq\mu_{2}\leq N-1
	\hspace{1 em} and \hspace{1 em} (\mu_{1}, \mu_{2})\neq (0,0)
	\end{displaymath}
is a solution of the {\it sl}(N;C) chiral field equation (\ref{3.4}). (Compare
with \cite{7}). Assume that our solution $\Theta=\Theta (\hbar; x,y,p,q)$ is 
analytic in $\hbar$. Consequently, from the results of Sec.2. one infers that
	\begin{equation}
	\label{3.12}
	\Theta_{0}=\Theta_{0} (0; x,y,p,q)
	\end{equation}
fulfils the H-P equation (\ref{2.1}).

Now if we put $\hbar=\frac{2\pi}{N}$ then $\hbar\rightarrow 0$ corresponds to
$N\rightarrow\infty$. Assume that the metric $ds^2$ given by (\ref{2.2}) for
$\Theta_{0}$ defined by (\ref{3.12}) describes the curved manifold. Thus one 
arrives at the sequence of {\it sl}(N;C) chiral fields for $N=2,3,\dots$ tending
to a curved heavenly space-time.

\section{Example}
\setcounter{equation}{0}
We look for the solution of the Moyal deformation of the H-P equation (\ref{2.3})
on $V\times T^2$, $V\subset R^2$, of the following form
	\begin{eqnarray}
	\label{4.1}	
	\partial_{x}\Theta & = & \exp(iq))\nonumber\\
	\partial_{y}\Theta & = & \sum_{m=0}^{\infty} 
	\eta_{(1,m)}(\hbar;y)\exp[i(p+mq)]	
	\end{eqnarray}
Substituting (\ref{4.1}) into (\ref{2.3}) employing also (\ref{2.8}) one gets 
the system of differential equations
	\begin{eqnarray}
	\label{4.2}
	\frac{d\eta_{(1,0)}}{dy} & = & 0\, ,\nonumber\\
	\frac{d\eta_{(1,m)}}{dy}-\frac{2}{\hbar}\sin(\frac{\hbar}{2})\, 
	\eta_{(1,m-1)}& = & 0\hspace{2 em}for\hspace{2 em}m\geq 1
	\end{eqnarray}
The general solution of the system (\ref{4.2}) reads
	\begin{equation}
	\label{4.3}
	\eta_{(1,m)}=\sum_{j=0}^{m}\frac{a_{m-j}}{j!}[\frac{2}{\hbar}
	\sin(\frac{\hbar}{2})\, y]^{j}\: ,
	\end{equation}
where $a_{0}, a_{1},\dots$ are complex constants (which may depend on $\hbar$).

Inserting (\ref{4.3}) into (\ref{4.1}) we find that the general 
$\partial_{y}\Theta$ of the form (\ref{4.1}) is any linear combination 
of the function $\exp(iq)\exp[\frac{2}{\hbar}\sin(\frac{\hbar}{2})\; y\exp(iq)]$
and all its derivatives with respect to y. 

For futher considerations we take the solution of Eq.(\ref{2.3}) to be
	\begin{eqnarray}
	\label{4.4}	
	\partial_{x}\Theta & = & \exp(iq))\nonumber\\
	\partial_{y}\Theta & = & 
	\exp(ip)\exp[\frac{2}{\hbar}\sin(\frac{\hbar}{2})\; y\exp(iq)]\nonumber\\
	& = & \sum_{m=0}^{\infty} 
	\frac{1}{m!}[\frac{2}{\hbar}\sin(\frac{\hbar}{2})\; y]^{m}\exp[i(p+mq)]	
	\end{eqnarray}
Hence
	\begin{equation}
	\label{4.5}
	\Theta=x\exp(iq)+\sum_{m=0}^{\infty} 
	\frac{(\frac{2}{\hbar}\sin\frac{\hbar}{2})^m}{(m+1)!}
	\, y^{m+1}\exp[i(p+mq)]
	\end{equation}
Evidently, the function $\Theta=\Theta (\hbar; x,y,p,q)$ defined by (\ref{4.5}) 
is analytic in $\hbar$ and also we can put $V=R^2$. Then
	\begin{eqnarray}
	\label{4.6}	
	\Theta_{0} & = & \Theta(0;x,y,p,q)=x\exp(iq)+\sum_{m=0}^{\infty} 
	\frac{1}{(m+1)!}\, y^{m+1}\exp[i(p+mq)]\nonumber\\
	& = & x\exp(iq)+\exp[i(p-q)]\{\exp[y\exp(iq)]-1\}
	\end{eqnarray}
is a solution of the Husain-Park heavenly equation (\ref{2.1}). Now we put in (4.4)
$\hbar=\frac{2\pi}{N}$. Then from (\ref{3.3}), (\ref{3.11}), (\ref{4.1}) and 
(\ref{4.4}) one gets
	\begin{eqnarray}
	\label{4.7}	
	 A_{x} & = & -\sum_{\mu =0}^{N-1}\{\sum_{r=0}^{\infty}
	\eta_{(1,\mu+Nr)}(\frac{2\pi}{N};y)\}L_{(1,\mu)} 
	\nonumber\\
	& = & -\sum_{\mu=0}^{N-1} \{ \sum_{r=0}^{\infty}
	\frac{1}{(\mu+Nr)!}[\frac{N}{\pi}\sin(\frac{\pi}{N})
	\; y]^{\mu+Nr}\} L_{(1 ,\mu )}  \nonumber\\
	 A_{y} & = & L_{(0,1)}
	\end{eqnarray}
As for $\mu < N	$
	\begin{equation}
	\label{4.8}
	\sum_{r=0}^{\infty}\frac{1}{(\mu+Nr)!}
	[\frac{N}{\pi}\sin(\frac{\pi}{N})\; y]^{\mu+Nr}=
	\frac{1}{N}\sum_{k =1}^{N}\omega^{(N-\mu )k}\exp[\omega^{k}
	\frac{N}{\pi}\sin(\frac{\pi}{N})\; y]
	\end{equation}
where $\omega=\exp\frac{2\pi i}{N}$ we obtain finally
	\begin{eqnarray}
	\label{4.9}
	A_{x} & = & -\sum_{\mu=0}^{N-1} \{\frac{1}{N}\sum_{k =1}^{N}
	\omega^{(N-\mu )k}\exp[\omega^{k}
	\frac{N}{\pi}\sin(\frac{\pi}{N})\; y]\} L_{(1 ,\mu )}
	\nonumber\\
	A_{y} & = & L_{(0,1)}
	\end{eqnarray}
Straightforward calculations show that the {\it sl}(N;C)-valued functions 
$A_{x}=A_{x}(x,y)$ and $A_{y}=A_{y}(x,y)$, indeed, fulfill the principal chiral model field equations 
(\ref{3.1}) and (\ref{3.2}) for every $N\geq 2$. Thus we get the sequence of the 
{\it sl}(N;C) chiral fields tending for $N\rightarrow\infty$ to the heavenly 
space-time of the metric defined by (\ref{2.2}) with the function $\Theta_{0}$
given by (\ref{4.6}). This metric can be written in the following null tetrad form
	\begin{eqnarray}
	\label{4.10}
	ds^2 & = & e^1\otimes e^2+e^2\otimes e^1+e^3\otimes e^4+e^4\otimes e^3
	\nonumber\\	
	e^1 & = & \Omega[idy-\exp(-iq)dp-ydq], \nonumber\\
	e^2 & = & \Omega[\exp(-iq)dp+ydq]\nonumber\\
	e^3 & = & \Omega dq , \nonumber\\ 
	e^4 & = & \Omega\exp[-(ip+y\exp(iq))]\{idx-\exp[-(ip+y\exp(iq))]dq\}
	\nonumber\\
	\Omega & = & \frac{1}{\sqrt{2}}\exp\{\frac{1}{2}[i(p+q)+y\exp(iq)]\}
	\end{eqnarray}
Then using the Cartan structure equations \cite{9,10} one finds that
the only nonzero null tetrad component of the Riemann curvature
tensor reads
	\begin{equation}
	\label{4.11}
	\frac{1}{2}C^{(1)}=R_{3131}=-2\Omega^{-2}
	[\frac{1}{4}\Omega^{-4}\exp(4iq)\, +1]
	\end{equation}
Consequently our heavenly space-time is of the type $[4]\times [-]$.
(see \cite{9}).

Note that if $x,y,p$ and $q$ are real coordinates then (\ref{4.10})
defines the complex heavenly metric on $R^2\times T^2$. Assuming 
$x,y,p$ and $q$ to be complex we get the complex heavenly space-time
(nonlinear graviton) of the holomorphic metric (\ref{4.10}).

\section*{Acknowledgments}
We are indebted to J.Tosiek and P.S{\l}oma for intrest in this work. 
One of us (F.J.T.) thanks Prof. C. Malinowska-Adamska and all the staff
of the Institute of Physics at Technical University of \L\'{o}d\'{z}, 
Poland, for warm hospitality. F.J.T. was supported by CINVESTAV and 
CONACyT, Mexico.


\begin{thebibliography}{20}
\bibitem{1}R.S.Ward,{\it Class. Quantum Grav.} \bf 7 \rm ,L 217 (1990).
\bibitem{2}Q.H.Park, {\it Int. J. Mod. Phys. } \bf A 7 \rm , 1415 (1992).
\bibitem{3}I.A.B.Strachan, {\it Phys.Lett. }\bf B 283\rm , 63 (1992)
\bibitem{4}V.Husain, {\it Class. Quantum Grav. }\bf 11 \rm , 927 (1994).
\bibitem{5}J.F.Pleba\'{n}ski, M.Przanowski and H.Garcia-Compe\'{a}n, {\it 
Mod. Phys. Lett. }\bf A 11 \rm , 663 (1996).
\bibitem{6}H.Garcia-Compe\'{a}n, J.F.Pleba\'{n}ski and M.Przanowski, 
{\it Phys. Lett. }\bf A 219 \rm , 249 (1996).
\bibitem{7}M.Przanowski and S.Forma\'{n}ski, "Searching for a universal integrable system"
to appear in {\it Acta Phys. Pol. }\bf B  \rm
\bibitem{8}D.B.Fairlie,P.Fletcher and C.K.Zachos {\it J. Math. Phys. }
\bf 31 \rm , 1088(1990)
\bibitem{9}J.F.Pleba\'{n}ski, {\it J. Math. Phys. } \bf 16 \rm , 2395 (1975).
\bibitem{10}J.F.Pleba\'{n}ski and M.Przanowski, {\it Acta Phys. Pol. }
 \bf B 19  \rm , 805 (1988).
	
		\end{thebibliography}
\end{document}